\def\bd{\begin{displaymath}}\def\ed{\end{displaymath}}
\def\be{\begin{equation}}\def\ee{\end{equation}}
\def\bea{\begin{eqnarray}}\def\eea{\end{eqnarray}}
\def\nn{\nonumber}\def\lb{\label}

\def\d{\delta}

\def\k{\kappa}\def\m{\mu}\def\o{\omega}\def\t{\tau}
\def\x{\xi}

\def\D{\Delta}\def\G{\Gamma}\def\L{\Lambda}

\def\de{\partial}
\def\mo{{-1}}\def\ha{{1\over 2}}

\def\ex{{\rm e}}

\def\eom{equations of motion }
\def\tran{transformation }\def\coo{coordinates }

\def\tl{transformation law }

\def\pb{Poisson brackets }

\def\poi{Poincar\'e }

\def\lt{Lorentz transformations }

\def\PL#1{Phys.\ Lett.\ {\bf#1}}
\def\PRL#1{Phys.\ Rev.\ Lett.\ {\bf#1}}
\def\PR#1{Phys.\ Rev.\ {\bf#1}}

 \def\IJMP#1{Int.\ J. Mod.\ Phys.\ {\bf #1}}
\def\MPL#1{Mod.\ Phys.\ Lett.\ {\bf #1}} 

\def\AoP#1{Ann.\ Phys.\ {\bf#1}}
\def\grq#1{{\tt gr-qc/\-#1}}\def\hep#1{{\tt hep-th/\-#1}}

\def\den{\left(1-{p_0\over\k}\right)}\def\pik{{p_1\over\k}}
\def\exs{\ex^{-2p_0/\k}}\def\trans{transformations }
\def\tls{transformation laws }

\documentclass[12pt]{article}
\begin{document}

\begin{titlepage}
\vspace{.3cm}
\begin{center}
\renewcommand{\thefootnote}{\fnsymbol{footnote}}
{\Large \bf Transformations of coordinates and}

{\Large \bf Hamiltonian formalism in deformed}

{\Large \bf Special Relativity}
\vfill%\vskip 15mm%27.mm
{\large \bf {S.~Mignemi\footnote{email: smignemi@unica.it}}}\\
\renewcommand{\thefootnote}{\arabic{footnote}}
\setcounter{footnote}{0}
\vfill%\vskip 7mm%1cm
{\small
  Dipartimento di Matematica, Universit\`a di Cagliari,\\
Viale Merello 92, 09123 Cagliari, Italy\\
\vspace*{0.4cm}
 INFN, Sezione di Cagliari\\
}
\end{center}
\vfill
\centerline{\bf Abstract}
\bigskip

We investigate the transformation laws of coordinates
in generalizations of special relativity with two
observer-independent scales.
The request of covariance leads to simple formulas if
one assumes noncanonical Poisson brackets, corresponding
to noncommuting spacetime coordinates.

\vfill
\end{titlepage}

\section{Introduction}
Recently, large interest has been devoted to a variety of
deformations of special relativity admitting two invariant
fundamental scales, the speed of light and an energy scale,
to be identified with Planck energy $\k$ [1-3].
These theories aim to describe the dynamics of particles up to the
Planck region, where the structure of spacetime may change due
to quantum gravity effects.
They were first introduced algebraically, investigating quantum
deformations of the Lorentz group \cite{LNR}, and later rederived
starting from a set of suitable physical postulates, as for
example the request that special relativity is recovered in the
low-energy limit \cite{AC}.

All these models have in common the assumption that the momentum of a
particle transforms nonlinearly under the Lorentz group, but
differ in several respects. In fact, first of all, it is possible to
construct many different nonlinear representations of the Lorentz group
satisfying the postulates mentioned above;
moreover, the theory can be defined either in standard
spacetime, or in a spacetime with noncommuting coordinates \cite{LRZ}.
In the latter case, when defining the classical dynamics through the
Hamiltonian formalism, one is forced to use noncanonical Poisson
brackets. Of course, depending on the choice of the previous  basic
assumptions, one obtains models that give rise to different physical
predictions.

In order to compare the theory with experiment, one also
needs a physical interpretation of the variables that enter in
the theory in terms of measurable quantities. This is not always
straightforward. We must consider in fact that the models under study,
although classical,
are supposed to be effective in the quantum domain, so that one cannot
expect that all quantities (as for example velocity) can be
defined in classical terms.

As we shall discuss in the following, for example, a difficulty in the
interpretation of the theory arises from the fact that the
transformation rules for the position of a particle depend on its
momentum, and hence particles occupying the same position in a
reference frame do not necessarily do so in another frame.

We believe that problems of this kind can be understod by recalling
the quantum-mechanical origin of the models, which only in the limit
of small momenta reduce to special relativity. This implies that the
physical quantities should be interpreted accordingly.
Clearly, we have no way to probe the sub-Planckian scales for which
the theory is defined, and so we do not have to wonder if some of the
usual assumption are not valid, provided that the theory is logically
consistent.

In this paper we try to clarify some of these questions by discussing
the issue of the transformation laws of position and momentum in a
Hamiltonian setting, assuming either commuting spacetime coordinates
and standard \pb or noncommuting \coo and deformed symplectic structure.
We treat explicitly the Magueijo-Smolin (MS) model \cite{MS},
since this is the simplest algebraically, but our considerations
are valid also for different models. In the appendix, we shortly
discuss the Lukierski-Nowicki-Ruegg (LNR) model \cite{LNR}.
For simplicity, we consider the case of 1+1 dimensions, but
our results can be easily extended to four dimensions.

\section{The transformation law for commuting coordinates}

Let us consider the Hamiltonian formulation for a free particle in
$1+1$ dimensions.
We introduce commuting spacetime coordinates $q^a$ and momenta $p_a$
($a=0,1$) for a particle, obeying canonical Poisson brackets, so that
\be
\{q^a,q^b\}=0,\qquad\{p_a,p_b\}=0,\qquad\{q^a,p_b\}=\d^a_b.
\ee
Deformed special relativity models [1-3] postulate a nonlinear
Lorentz transformation law for the momenta, but do not specify
the \tl of the coordinates. More precisely, it is assumed that
under an infinitesimal boost of generator $J$,
\be\lb{tl}
\d p_a=\{J,p_a\}=w_a(p),
\ee
where $w_a(p)$ is a nonlinear function of the momentum $p_a$.
For the MS model \cite{MS}, the functions $w_a$ are
\be\lb{inptr}
w_0=p_1-{p_0p_1\over\k},\qquad
w_1=p_0-{p_1^2\over\k}.
\ee

Now, it is easy to see that the \tl (\ref{tl}) implies that also
the \coo must transform in a nontrivial way under Lorentz
transformations. This is a simple consequence of the Jacobi
identities. In fact, under a \lt $\d q^a=\{J,q^a\}$, but
\be
\{\{J,q^a\},p_b\}+\{\{p_b,J\},q^a\}+\{\{q^a,p_b\},J\}=0
\ee
and hence
\be
{\de\{J,q^a\}\over\de q^b}=-\{\{J,q^a\},p_b\}=-{\de w_b\over\de p_a}
\ee
Integrating these relations, one finds that
\be\lb{ict}
\{J,q^a\}=-{\de w_b\over\de p_a}q^b+f^a(p)
\ee
where $f^a(p)$ are arbitrary functions of $p^a$, satisfying
${\de f^a/\de p_b}={\de f^b/\de p_a}$,  that can be set to
zero.

From (\ref{inptr}) and (\ref{ict}) one has
\be\lb{inqtr}
\d q^0=-q^1+{p_1\over\k}q^0,\qquad \d q^1=-\left(1-{p_0\over\k}\right)
q^0+{2p_1\over\k}q^1,
\ee

Hence, the \tran properties of the position of a particle depend
on its momentum, and two particles occupying the same position in
a reference frame can occupy  different positions in another.
The momentum dependence of the transformations of the coordinates
of a particle has been first remarked by Kowalski-Glikman \cite{KG}
in a special case, but it seems to have been disregarded by other
authors.

Actually, one can even derive the finite transformation properties
of coordinates from those of momenta, by requiring covariance.
In momentum space $\cal P$, the effect of a boost is given by a
nonlinear \tran
\be\lb{p'}
p_a\to p'_a=W_a(p).
\ee
The space of \coo can be
identified with the tangent space to $\cal P$ and hence for
covariance,
\be\lb{q'}
q^a\to q'^a=\L^a_{\ b}\,q^b,
\ee
where
\be\lb{Lambda}
\L^a_{\ b}=\left(\de W_a\over\de p_b\right)^\mo.
\ee

For the MS model, one has \cite{MS}
\be\lb{W}
W_0={p_0\cosh\x+p_1\sinh\x\over\D},\qquad
W_1={p_1\cosh\x+p_0\sinh\x\over\D},
\ee
where
\bd
\D=1+{p_0\over\k}(\cosh\x-1)+{p_1\over\k}\sinh\x,
\ed
and $\x$ is the rapidity parameter.
It follows that
\be\lb{lambda}
\L^a_{\ b}= \D\left(
\begin{array}{cc}
\cosh\x+{p_0\over\k}(1-\cosh\x)&-\sinh\x+{p_1\over\k}(1-\cosh\x)\\
-\sinh\x+{p_0\over\k}\sinh\x&\cosh\x+{p_1\over\k}\sinh\x
\end{array}
\right)
\ee
It is easy to check that the infinitesimal version of these relations is
(\ref{inqtr}).

Given the \tl (\ref{q'}), one can also define a covariant lagrangian,
\be
L=p_a\dot q^a-H(p).
\ee
In fact, $p_a\dot q^a$ is clearly invariant up to a total derivative
under the \lt (\ref{p'}-\ref{q'}).
For a free particle, the Hamiltonian $H$ is given by the Casimir
invariant of the algebra \cite{MS},
\be\lb{H}
H=\ha\,{p_0^2-p_1^2\over\den^2}.
\ee
It may be interesting to write down the Hamilton equations
for (\ref{H}):
\bd
\dot q^0={\de H\over\de p_0}={p_0-{p_1^2\over\k}\over\den^3},\qquad
\dot q^1={\de H\over\de p_1}=-{p_1\over\den^2},
\ed
\bd
\dot p_0=-{\de H\over\de q^0}=0,\qquad
\dot p_1=-{\de H\over\de q^1}=0.
\ed
The velocity of a particle obeying these \eom is given by
\be
v={\dot q_1\over\dot q_0}=
{p_1-p_0p_1/\k\over p_0-p_1^2/\k}.
\ee

As discussed in the literature \cite{KM,Mi}, this definition of
velocity is not satisfactory, since it implies
that the velocity of a particle depends on its mass, and it
is difficult to reconcile this fact with the role of velocity as
the parameter of the Lorentz transformations.

A solution
to this problem is to use deformed Poisson brackets \cite{LRZ}. This
is very natural from the point of view of $\k$-\poi models, since
deformed \pb can be considered as the classical limit of noncommuting
spacetime coordinates. In the following section we reconsider the
MS model from this point of view.

\section{The transformation law for noncommuting coordinates }

In the MS model, a suitable definition of the velocity of a particle
is given by $v=p_1/p_0$ \cite{Gr}.
This coincides with $\dot q_1/\dot q_0$
if one introduces the following symplectic structure \cite{Gr}:
\bea\lb{sym}
&&\{q^0,q^1\}={q^1\over\k},\quad\{p_0,p_1\}=0,\quad\{q^0,p_0\}=
1-{p_0\over\k},\cr
&&\{q^1,p_1\}=1,\quad\{q^0,p_1\}=-{p_1\over\k},\qquad\{q^1,p_0\}=0.
\eea
The infinitesimal transformations of \coo can then be deduced
from the Jacobi identities as above. They read
\be
\d q^0=-q^1+{p_1\over\k}q^0,\qquad\d q^1=-q^0+{p_1\over\k}q^1.
\ee
For a free particle, with Hamiltonian (\ref{H}),
the Hamilton equations derived from (\ref{sym}) are
\bea
\dot q^0&=&\den{\de H\over\de p_0}-\pik{\de H\over\de p_1}=
{p_0\over\den^2},\cr
\dot q^1&=&{\de H\over\de p_1}=-{p_1\over\den^2},\cr
\dot p_0&=&-\den{\de H\over\de q^0}=0,\cr
\dot p_1&=&-{\de H\over\de q^1}+\pik{\de H\over\de q^0}=0,\nn
\eea
and hence $v=p_1/p_0$, as expected.

It is known that
the Hamilton equations for systems with nonstandard symplectic
structure can be derived from an action principle \cite{Sa}.
Given a phase space with symplectic structure
$\{Q^A,Q^B\}=\o^{AB}$, where $Q^A$ denote either the \coo or
the momenta, one defines the functions $R_A(Q^A)$ such that
\be
{\de R_A\over \de Q^B}-{\de R_B\over \de Q^A}=\o_{AB},
\ee
where $\o_{AB}$ is the inverse of $\o^{AB}$.
The Hamilton equations can then be obtained varying with
respect to $Q^A$ the action
\be
I=\int d\t(R_A\dot Q^A-H).
\ee
Note that in general the action so defined contains derivatives
of the momenta.

In our case, this procedure yields
\be\lb{ac}
I=\int d\t\left[-\k\log\den\,\dot q^0+p_1\dot q^1-
{p_1q^1\over\k\den}\,\dot p_0-H\right].
\ee
After integration by parts, (\ref{ac}) can also be written as
\be\lb{act}
I=-\int d\t\left[{q^0+{p_1\over\k}q^1\over\den}\,\dot p_0
+q^1\dot p_1+H\right].
\ee

The \tls for the \coo can now be obtained by requiring that the
variables $r^a$ conjugate to the momenta $p_a$ in (\ref{act})
transform covariantly,
i.e., according to $r^a\to r'^a=\L^a_{\ b}r^b$. From (\ref{act})
one has
\be
q^0=\den r^0-{p_1\over\k}r^1,\qquad q^1=r^1,
\ee
and hence
\bd
q'^0=\left(1-{W_0\over\k}\right)\L^0_{\ a}r^a-{W_1\over\k}\L^1_{\ a}r^a,
\qquad q'^1=\L^1_{\ a}r^a.
\ed
Substituting (\ref{W}) and (\ref{lambda}), after tedious but elementary
algebraic manipulations, one gets the surprisingly simple result
\bea\lb{nct}
&&q'^0=\D(q^0\cosh\x-q^1\sinh\x),\cr
&&q'^1=\D(-q^0\sinh\x+q^1\cosh\x).
\eea

Thus, the \tls of \coo are identical to the usual Lorentz
transformations, except for the momentum-dependent factor $\D$.
It is then easy to build a simple ``invariant length"
\be
(p_ap^a)\ q^aq_a\sim\den^2\ q^aq_a,
\ee
from which an invariant line element can be defined.
One may also define new (commuting) \coo
$\bar q^a=(1-p_0/\k)q^a$, which transform according to standard
Lorentz transformations, so that $\bar q^a\bar q_a$ is invariant.
These \coo however do not lead to a correct definition of the
velocity of a particle.

\section{Conclusions}
We have discussed the coordinate \tls that allow a
covariant definition of the Hamiltonian formalism in models of
deformed special relativity, both in the case of commuting \coo
with canonical symplectic structure, and of noncommuting \coo
with deformed Poisson brackets.
The \tls for \coo are momentum dependent and in general rather
complicated, except in the case of the MS model with noncommuting
coordinates. In this case the \tls assume a natural
form, that allows the definition of a simple (momentum-dependent)
line element.

Of course, the interpretation of the momentum dependence of the
\tls for \coo is highly nontrivial. The deformed Lorentz \trans
act on the full phase space and not separately on coordinate and
momentum space. This implies that the coincidence of two events
becomes observer-dependent. At first, this prediction may seem
unphysical, but it must be considered that the theory is assumed
to be effective at sub-Planckian scales, of which we do not have
any direct experience.

A more conservative interpretation of the results of this paper
would be to consider the $q^a$ simply
as labels of the position of a particle, which should be connected
to the spacetime coordinates $x^\m$ by a sort of vierbein field:
$q^a=e^a_\m x^\m$. The consistency of this approach is currently
being investigated.

These considerations are also interesting in view of the inclusion
of gravity in the theory \cite{Mi'}. A consistent approach should
presumably lead to some kind of phase space extension of general
relativity.

{\bf Note.} While completing this work, I became aware of a paper
\cite{KMM} where the same transformations of \coo (\ref{nct}) are
proposed, starting from a different point of view.

\section{Appendix}
We report here the calculations analogous to those presented
above,
for the case of the LNR model \cite{LNR}.

The infinitesimal Lorentz \trans for LNR are given by
(\ref{tl}), with
\be
w_0=p_1,\qquad
w_1={\k\over2}\left(1-\exs-{p_1^2\over\k^2}\right).
\ee
It follows from (\ref{ict}) that
\be
\d q^0=\exs\,q^1,\qquad\d q^1=q^0-{p_1\over\k}q^1.
\ee

The finite \trans are given by (\ref{p'}), with \cite{BAK}
\be
W_0=p_0+\k\log\G,\qquad
W_1={p_1\cosh\x+{\k\over2}\left(1-\exs+{p_1^2\over\k^2}\right)
\sinh\x\over\G},
\ee
where
\bd
\G=\ha\left(1+\exs-{p_1^2\over\k^2}\right)+\ha\left(1-\exs+
{p_1^2\over\k^2}\right)\cosh\x+{p_1\over\k}\sinh\x.
\ed
From (\ref{Lambda}) follows that $q'^a=\L^a_{\ b}q^b$, with
\bea
\L^1_{\ 1}&=&\ha\left(1-\exs-{p_1^2\over\k^2}\right)+\ha
\left(1+\exs+{p_1^2\over\k^2}\right)\cosh\x+{p_1\over\k}\sinh\x,\cr
\L^1_{\ 0}&=&\exs\left[{p_1\over\k}\left(1-\cosh\x\right)-\sinh\x\right],
\eea
and $\L^0_{\ 0}=\L^1_{\ 1}/\G$, $\L^1_{\ 0}=\L^0_{\ 1}/\G$.

Also in the LNR case the standard hamiltonian formalism does not give
a consistent definition of the particle velocity.
A suitable definition of velocity is given instead by the
right-invariant velocity \cite{THM},
\be\lb{vel}
v={\ex^{p_0/\k}p_1\over\k\sinh{p_0\over\k}+\ex^{p_0/\k}{p_1^2\over\k}},
\ee
and can be obtained introducing the following nonstandard symplectic
structure \cite{LN}:
\bea
&&\{q^0,q^1\}={q^1\over\k},\quad\{p_0,p_1\}=0,\quad\{q^0,p_0\}=1,\cr
&&\{q^1,p_1\}=1,\quad\{q^0,p_1\}=-{p_1\over\k},\qquad\{q^1,p_0\}=0.
\eea
The infinitesimal \trans of \coo can then be obtained as before from
the Jacobi identities. They read
\be
\d q^0=-{\k\over2}\left(1-\exs-{p_1^2\over\k^2}\right)q^1+{p_1\over\k}q^0,
\qquad\d q^1=-q^0.
\ee

The Hamiltonian for a free particle is given by the Casimir invariant
\be
H=\ha\left[\left(2\k\sinh{p_0\over2\k}\right)^2-\ex^{p_0/\k}
{p_1^2\over\k^2}\right],
\ee
and the Hamilton equations are
\bea\lb{he}
\dot q^0&=&{\de H\over\de p_0}-\pik{\de H\over\de p_1}=
\k\sinh{p_0\over\k}+\ex^{p_0/\k}{p_1^2\over\k},\cr
\dot q^1&=&{\de H\over\de p_1}=-\ex^{p_0/\k}p_1,\cr
\dot p_0&=&-{\de H\over\de q^0}=0,\cr
\dot p_1&=&-{\de H\over\de q^1}+\pik{\de H\over\de q^0}=0,
\eea
which lead to the definition (\ref{vel}) of velocity.

As explained above, eqs. (\ref{he}) can be obtained from an action
principle. The action reads in this case,
\be
I=\int d\t\left[p_0\dot q^0+p_1\dot q^1-{q^1p_1\over\k}\,\dot p_0
-H\right],
\ee
or, after integration by parts,
\be\lb{act'}
I=-\int d\t\left[\left(q^0+{p_1\over\k}q^1\right)\dot p_0
+q^1\dot p_1+H\right].
\ee

The \tls for the \coo can now be obtained by requiring that the
variables $r^a$ conjugate to the momenta $p_a$ in (\ref{act'})
transform covariantly. One has
\be
q^0=r^0-{p_1\over\k}r^1,\qquad q^1=r^1,
\ee
and hence
\bd
q'^0=\L^0_{\ a}r^a-{1\over\k}W_1\L^1_{\ a}r^a,\qquad q'^1=\L^1_{\ a}r^a.
\ed
After some algebra, one gets the result
\bea
q'^0&=&q^0(\cosh\x+\pik\sinh\x)
-{q^1\over2}\left(1+\exs-{p_1^2\over\k^2}\right)\sinh\x,\cr
q'^1&=&-q^0\left[\sinh\x+\pik(1-\cosh\x)\right]\cr
&&+{q^1\over2}\left[\left(1-\exs+{p_1^2\over\k^2}\right)+\left(1+
\exs-{p_1^2\over\k^2}\right)\cosh\x\right].\nn
\eea
In this case, the \tls for \coo are not especially simple and do not
seem to lead to the same interesting developments as in the MS case.

\end{document}